\documentclass[twocolumn,superscriptaddress,nofootinbib]{revtex4}
\usepackage{graphicx}
\usepackage{amsmath,amsfonts}
\usepackage{amssymb,arydshln,array,dcolumn}
\usepackage{hyperref}

\usepackage{fancyhdr}

\newcommand{\be}{\begin{equation}}
\newcommand{\ee}{\end{equation}}
\newcommand{\beq}{\begin{equation}}
\newcommand{\eeq}{\end{equation}}
\newcommand{\ba}{\begin{array}}
\newcommand{\ea}{\end{array}}
\newcommand{\bea}{\begin{eqnarray}}
\newcommand{\eea}{\end{eqnarray}}
\newcommand{\bean}{\begin{eqnarray*}}
\newcommand{\eean}{\end{eqnarray*}}

\newcommand{\IZ}{\mathbb{Z}}

\newcommand{\cN}{{\cal N}}

\newcommand{\cA}{{\cal A}}
\newcommand{\cB}{{\cal B}}
\newcommand{\cC}{{\cal C}}

\newcommand{\cV}{{\cal V}}

\def\cjn1{{\cA, \cC^*\otimes \wedge^j \cN^*}}
\def\bjn1{{\cA, \cB^*\otimes \wedge^j \cN^*}}
\def\vjn1{{\cA, \cV^*\otimes \wedge^j \cN^*}}
\def\cjn2{{\cA, \cC\otimes \wedge^j \cN^*}}
\def\bjn2{{\cA, \cB\otimes \wedge^j \cN^*}}
\def\vjn2{{\cA, \cV\otimes \wedge^j \cN^*}}

\begin{document}



\title{Non-generic Couplings in Supersymmetric Standard Models}
\author{Evgeny I. Buchbinder}
\affiliation{The University of Western Australia,
35 Stirling Highway, Crawley WA 6009, Australia}
\author{Andrei Constantin}
\affiliation{Rudolf Peierls Centre for Theoretical Physics, 
Oxford University, 1 Keble Road, Oxford, OX1 3NP, U.K.}
\author{Andre Lukas}
\affiliation{Rudolf Peierls Centre for Theoretical Physics, 
Oxford University, 1 Keble Road, Oxford, OX1 3NP, U.K.}

\date{\today}

\pagestyle{fancy}
\fancyhf{}
\fancyhead[L]{Buchbinder, Constantin, Lukas}
\fancyhead[R]{ Non-generic Couplings}

\begin{abstract}
We study two phases of a heterotic standard model, obtained from a Calabi-Yau compactification of the $E_8\times E_8$ heterotic string, in the context of the associated four-dimensional effective theories. In the first phase we have a standard model gauge group, an MSSM spectrum, four additional $U(1)$ symmetries and singlet fields. In the second phase, obtained from the first by continuing along the singlet directions, three of the additional $U(1)$ symmetries are spontaneously broken and the remaining one is a $B-L$ symmetry. In this second phase, dimension five operators inducing proton decay are consistent with all symmetries and as such, they are expected to be present. We show that, contrary to this expectation, these operators are forbidden due to the additional $U(1)$ symmetries present in the first phase of the model. We emphasize that such ``unexpected" absences of operators, due to symmetry enhancement at specific loci in the moduli space, can be phenomenologically relevant and, in the present case, protect the model from fast proton decay.
\end{abstract}


\maketitle

\section{Introduction}

A widely accepted dictum is that all the couplings that are allowed by the symmetries of an effective field theory (EFT) should be present in the Lagrangian. In the present letter, we point out that care has to be taken when this principle is applied to EFTs from string theory. We will present an explicit example, in the context of a standard model derived from heterotic string theory on Calabi-Yau manifolds, where this principle appears to be violated, at least when thinking about the associated EFT in the standard way. One of the relevant key facts is that string theory can lead to symmetry enhancement at particular loci in moduli space. These additional symmetries are not directly visible at a generic locus since the corresponding gauge bosons are massive and removed from the EFT. Yet, these symmetries can still forbid certain operators everywhere in moduli space, thereby leading to ``unexpected" absences of operators.

Our example model is based on a heterotic line bundle model on a certain Calabi-Yau manifold which has been constructed in two previous publications \cite{Buchbinder:2013dna, Buchbinder:2014qda}. Here, we will focus on the associated low-energy theory and explain the effect purely in terms of the four-dimensional $ N=1$ EFT. We will discuss and compare two phases of this EFT, both of which have been obtained from a string construction. The first phase arises at a specific locus in moduli space and corresponds to an MSSM with four additional $U(1)$ symmetries and a number of fields uncharged under the Standard Model gauge group. The additional singlets in this model can be interpreted as bundle moduli and, from a low-energy perspective, they are candidates for right-handed neutrinos. The second phase which arises at a more generic locus in moduli space corresponds to an MSSM with an additional $U_{B-L}(1)$ symmetry. In low-energy terms, it can be obtained from the first phase by continuation along the singlet directions thereby spontaneously breaking three of the four $U(1)$ symmetries while leaving $U_{B-L}(1)$ unbroken. 

Our point concerns the allowed operators in the second, generic phase with $U_{B-L}(1)$ symmetry.\footnote{The $U_{B-L}(1)$ symmetry 
is a linear combination of the hypercharge and an additional $U(1)$ symmetry with massive gauge boson. The latter $U(1)$ manifests itself at low energies as a {\it global} symmetry. This approach is different from the one studied in~\cite{Braun:2005ux, Braun:2005bw}. To prevent the proton from fast decay the authors in~\cite{Braun:2005ux, Braun:2005bw} considered models with {\it local}
$U_{B-L}(1)$ symmetry which then has to be violated by radiative corrections at scales below the string scale but higher 
than the electroweak scale~\cite{Ambroso:2009jd, Ambroso:2009sc}.} 
It is well-known that dimension five operators inducing proton decay are allowed by $U_{B-L}(1)$. Following the general lore, we should, therefore, expect that these operators are present in the generic phase of our model. This would imply a serious phenomenological problem with proton stability. However, it turns out that the enhanced $U(1)^4$ gauge symmetry which arises at the specific locus in moduli space comes to the rescue. Not only does this enhanced symmetry forbid the dimension five operators in question, it also forbids all such operators with additional singlet insertions. This means that these operators remain forbidden even when we turn on singlet vacuum expectation values and move away from the enhanced symmetry locus. 

To explain this in detail, we define the two relevant effective field theories in Sections~\ref{sec:AbelianLocus} and~\ref{sec:NonAbelianLocus}, respectively, and present their field content and their allowed superpotential couplings. We discuss the implications of the $U(1)$~symmetries at the enhanced symmetry locus and throughout the moduli space, in particular the resulting 
absence of dimension five proton decay operators. We conclude in Section~\ref{sec:Conclusion}.



\section{The theory with enhanced symmetry}\label{sec:AbelianLocus}


We begin by describing the four-dimensional $N=1$ theory at the locus with enhanced symmetry, starting with the particle spectrum and followed by the key features of the effective action.

\subsection{Spectrum}
We consider an effective field theory with the standard model gauge group $G_{\text{SM}}= SU(3) \times SU(2) \times U(1)$
and with an additional $U(1)^4$ symmetry group which significantly constrains the theory. 
Such models with extra $U(1)$  symmetries arise from compactifications 
of the $E_8 \times E_8$ heterotic string theory at specific loci in the moduli space where the structure group of the vector bundle 
degenerates~\cite{Anderson:2009sw, Anderson:2009nt, Anderson:2012yf}. The gauge bosons of these extra $U(1)$ groups 
can be massive or massless depending on the details of the model. If the gauge boson 
is massive the corresponding $U(1)$ group appears at low energies as a global symmetry.
It is convenient to describe these additional $U(1)$ symmetries by the group $S(U(1)^5)$ whose factors we label by indices $a,b,\dots =1,\ldots 5$. Its representations are denoted by five-dimensional integral vectors
\begin{equation}
\mathbf{q} = \left( q_1,\ldots, q_5\right)
\end{equation}
with the understanding that two charge vectors $\mathbf{q}$ and $\mathbf{q'}$ are identified, if $\mathbf{q}-\mathbf{q'} \in \IZ\mathbf{n}$, where $\mathbf{n} = \left(1,1,1,1,1\right)$.

The gravitational spectrum of the model consists of the dilaton, $S$, four K\"ahler moduli $T^i=t^i+i\chi^i$ (where $t^i$ are the geometrical fields, measuring the size of Calabi-Yau two-cycles, and $\chi^i$ are the associated axions) plus complex structure moduli which will not play an essential role in our discussion. The axions $\chi^i$ transform non-linearly under the $S(U(1)^5)$ symmetry$\,$\footnote{The dilatonic axion also receives a non-trivial transformation at one-loop order. However, this does not affect our discussion.} as
\begin{equation}
 \delta_a\chi^i=-{k_a}^i\; . \label{chishifts}
\end{equation}

In the rest of the paper we will concentrate on the specific model constructed in~\cite{Buchbinder:2013dna, Buchbinder:2014qda}. 
For our model, the integers ${k_a}^i$ are explicitly given by
\begin{equation}\label{kmat}
({k_a}^i)=
\left(
\begin{array}{ccccc}
 -1 & -1 & ~~0 & ~~1 & ~~1~ \\
~~0 & -3 & ~~1 & ~~1 & ~~1~ \\
~~0 & ~~2 & -1 & -1 & ~~0 ~\\
 ~~1 & ~~2 &~~ 0 & -1 & -2 ~\\
\end{array}
\right)
\end{equation}

Let us review the properties of the resulting low--energy theory (see~\cite{Buchbinder:2013dna, Buchbinder:2014qda} for details). 
As was discussed above, the symmetry group of the low-energy effective theory is $G_{\text{SM}}\times S(U(1)^5)$. 
In this case, three out of the four $U(1)$~gauge bosons receive string scale St\"uckelberg masses and the remaining one is massless. 

The matter spectrum consists of the following multiplets
\begin{equation}\label{spectrum}
\begin{array}{llllll}
 2\,Q_{\mathbf{e}_2}~~& 2\,u_{\mathbf{e}_2}~~~~~& 2\,e_{\mathbf{e}_2}~~~ ~~& Q_{\mathbf{e}_4}~~~~~& u_{\mathbf{e}_4}~~~~~& e_{\mathbf{e}_4} \\[2pt]
 &2\,L_{\mathbf{e}_4+\mathbf{e}_5} & 2\,d_{\mathbf{e}_4+\mathbf{e}_5} & L_{\mathbf{e}_2+\mathbf{e}_5}&d_{\mathbf{e}_2+\mathbf{e}_5}\\[2pt]
 &H_{\mathbf{e}_2+\mathbf{e}_4}& \bar H_{-\mathbf{e}_2-\mathbf{e}_4}\\[5pt]
 3\,S_{\mathbf{e}_2-\mathbf{e}_1} & 3\,S_{\mathbf{e}_4-\mathbf{e}_1}&5\,S_{\mathbf{e}_2-\mathbf{e}_3} & 3\,S_{\mathbf{e}_2-\mathbf{e}_5} & S_{\mathbf{e}_4-\mathbf{e}_3}\; ,
\end{array}
\end{equation}
where the subscripts indicate the $S(U(1)^5)$ charges and $\mathbf{e}_a$ denote the standard unit vectors in five dimensions. The first three lines represent a perfect MSSM spectrum, however with specific $S(U(1)^5)$ charges for each multiplet. In addition, we also have a spectrum of singlets, $S$, which are neutral under the standard model group but charged under $S(U(1)^5)$. Note that the $S(U(1)^5)$ charge of the standard model multiplets only depends on the $SU(5)$ GUT multiplet they reside in, so that, for each family, the multiplets in $\mathbf{10}=\left[Q,u,e\right]$ have the same $S(U(1)^5)$ charge, as do the multiplets in $\mathbf{\overline{5}}=\left[d,L\right]$. This fact is related to the underlying group structure of the model, which originates from an $SU(5)$ GUT broken by a Wilson line. 

The above spectrum is apparently anomalous. Indeed, one can compute the mixed $U(1)\,$-$\,G_{\text{SM}}^2$ anomaly to find
\begin{equation}
\cA_{U(1)-G_{\text{SM}}^2} = \sum_{\text{all families}}\left( 3\,{\mathbf q}(\mathbf{10}) + {\mathbf q}(\mathbf{\overline{5}})\right) = (0,7,0,5,3)\; .
\end{equation}
However, these anomalies (as well as the cubic and mixed gravitational anomalies) are cancelled by the Green-Schwarz mechanism, facilitated by the axionic shifts~\eqref{chishifts}.

If we describe linear combinations of the $U(1)$ symmetries by vectors ${\bf v}=(v^a)$ (demanding that ${\bf v}\cdot {\bf n}=0$ to remove the overall $U(1)$), then massless vector bosons are characterized by the equation ${k_a}^iv^a=0$. Applying this to Eq.~\eqref{kmat} shows that, for our model, three of the four $U(1)$ symmetries are St\"uckelberg massive, while the linear combination ${\bf v}=(-4, 1,6, -4, 1)$ remains massless.

\subsection{Effective action}
The K\"ahler potential has the standard form
\begin{equation}
 K=-\log(S+\bar{S})-\log(\kappa)+K_{\rm cs}+G_{IJ}C^I\bar{C}^J\; ,
 \end{equation}
where $K_{\rm cs}$ is the complex structure K\"ahler potential which will not be needed explicitly and $C^I$ collectively denote all matter fields listed previously. The specific form of the matter field K\"ahler metric $G_{IJ}$ is not relevant to our discussion and it will be sufficient to know that it is positive definite. The pre-potential, $\kappa$, for the K\"ahler moduli is explicitly given by$\,$\footnote{For ease of notation, we will write explicit indices of the fields $t^i$ as subscripts.}
\begin{equation}
 \kappa=d_{ijk}t^it^jt^k=12\left( t_1\,t_2\,t_3+t_1\,t_2\,t_4+t_1\,t_3\,t_4+t_2\,t_3\,t_4\right)\; ,
\end{equation}
and this equation defines the topological numbers $d_{ijk}$ for our model. We also note that the allowed range of the moduli $t^i$ (the K\"ahler cone of the underlying manifold) is $t^i>0$, for $i=1,2,3,4$. 

From this K\"ahler potential and the $S(U(1)^5)$ symmetry transformations given earlier, we can compute the $S(U(1)^5)$ D-terms $D_a$. 
Their general form is~\cite{Anderson:2009nt}$\,$\footnote{In addition, there is also a one-loop correction to this D-term, resulting from the transformation of the dilatonic axion, which we omit. This correction does not affect our discussion.}
\begin{equation}
 D_a=\frac{3}{\kappa}\, {k_a}^id_{ijk} t^jt^k+\sum_{I,J}q_a(C_I)C^I\bar{C}^J
\end{equation}
where $q_a(C_I)$ denotes the $S(U(1)^5)$ charges of the matter fields. Due to the special unitary nature of the group these D-terms satisfy the constraint $\sum_{a=1}^5 D_a=0$. The first term corresponds to a Fayet-Illiopoulos (FI) term which is explicitly given by
\beq\label{eq:Dterm20}
\frac{12}{\kappa}\ 
\left(
\begin{array}{c}
t_1 t_2 + t_1 t_3 - t_2 t_4 - t_3 t_4 \\
4\, t_1 t_2 - t_1 t_3 + t_2 t_3 - t_1 t_4 + t_2 t_4 - 4\, t_3 t_4\\
-t_1 t_2  + t_1 t_3 - t_2 t_4 + t_3 t_4\\
-2\, t_1 t_2 + 2\, t_3 t_4\\\
-2\,t_1 t_2 - t_1 t_3 - t_2 t_3 + t_1 t_4 + t_2 t_4 + 2\, t_3 t_4
\end{array}
\right) 
\eeq
while the second term is the matter field contribution which reads~\footnote{For simplicity, we omit the matter field K\"ahler metric $G_{IJ}$. Since $G_{IJ}$ is positive definite this will not affect our conclusions.}
\beq\label{eq:Dterm1}
\left(
\begin{array}{c}
-S_{\mathbf{e}_2-\mathbf{e}_1} S^{\dagger}{}_{\mathbf{e}_2-\mathbf{e}_1} - S_{\mathbf{e}_4-\mathbf{e}_1} S^{\dagger}{}_{\mathbf{e}_4-\mathbf{e}_1}\\
S_{\mathbf{e}_2-\mathbf{e}_1} S^{\dagger}{}_{\mathbf{e}_2-\mathbf{e}_1} + S_{\mathbf{e}_2-\mathbf{e}_3} S^{\dagger}{}_{\mathbf{e}_2-\mathbf{e}_3} + S_{\mathbf{e}_2-\mathbf{e}_5} S^{\dagger}{}_{\mathbf{e}_2-\mathbf{e}_5}\\
-S_{\mathbf{e}_2-\mathbf{e}_3} S^{\dagger}{}_{\mathbf{e}_2-\mathbf{e}_3} - S_{\mathbf{e}_4-\mathbf{e}_3} S^{\dagger}{}_{\mathbf{e}_4-\mathbf{e}_3}\\
-S_{\mathbf{e}_2-\mathbf{e}_5} S^{\dagger}{}_{\mathbf{e}_2-\mathbf{e}_5}\\
S_{\mathbf{e}_4-\mathbf{e}_1} S^{\dagger}{}_{\mathbf{e}_4-\mathbf{e}_1} + S_{\mathbf{e}_4-\mathbf{e}_3} S^{\dagger}{}_{\mathbf{e}_4-\mathbf{e}_3}
\end{array}
\right)
\eeq
Finally, the superpotential is severely restricted by the $S(U(1)^5$ symmetry and only contains the terms
\begin{equation}\label{eq:W}
\begin{aligned}
W=\lambda_i\bar{H}_{-\mathbf{e}_2-\mathbf{e}_4}& \left(Q_{\mathbf{e}_2}^{(i)}u_{\mathbf{e}_4}+Q_{\mathbf{e}_4}u_{\mathbf{e}_2}^{(i)}\right)\\
&+\rho_{\alpha i}\,S_{\mathbf{e}_2-\mathbf{e}_5}^{(\alpha)}L_{\mathbf{e}_4+\mathbf{e}_5}^{(i)}\bar{H}_{-\mathbf{e}_2-\mathbf{e}_4}\; .
\end{aligned}
\end{equation} 
Here, $i=1,2$ labels the two families with the same $S(U(1)^5)$ charges  and $\alpha=1,2,3$ labels the three singlets $S_{\mathbf{e}_2-\mathbf{e}_4}$. We emphasise that these are all the allowed superpotential terms, including possible higher-dimensional operators with or without singlet insertions. In particular, we note the absence of any dimension four and five operators which can induce proton decay. 
Additionally, we see that the down-Yukawa matrix vanishes perturbatively~\footnote{A non-zero down-Yukawa matrix may be generated by non-perturbative effects.} and the up-Yukawa matrix has rank 2. The actual quark masses depend on the proper normalization of the kinetic terms in the action which is beyond the scope of the present paper and will not be relevant to the main point we would like to make.

We should now study the supersymmetric moduli space of this model, taking into account the K\"ahler moduli $t^i$ and the singlet fields $S$ . Since the above superpotential has no pure singlet field part, this amounts to studying the D-flat directions of the model. We begin with the specific locus where all singlet field VEVs vanish, $\langle S\rangle =0$, and where the additional $U(1)$ symmetries are not spontaneously broken (although three of them are St\"uckelberg heavy). To satisfy the D-flat conditions in this case, the FI terms have to be set to zero individually which is equivalent to $t_1=t_2=t_3=t_4$. 

If we move away from this specific locus in a generic way, by switching on all singlet VEVs, the non-zero matter field parts of the D-terms can be compensated for by the FI terms with suitable choices of the K\"ahler moduli, provided they satisfy the inequalities
\beq\label{eq:Dterm2}
\begin{array}{c}
t_1 t_2 + t_1 t_3 - t_2 t_4 - t_3 t_4 \geq 0\\
4\, t_1 t_2 - t_1 t_3 + t_2 t_3 - t_1 t_4 + t_2 t_4 - 4\, t_3 t_4 \leq 0\\
-t_1 t_2  + t_1 t_3 - t_2 t_4 + t_3 t_4 \geq0\\
-2\, t_1 t_2 + 2\, t_3 t_4 \leq 0\\
-2\,t_1 t_2 - t_1 t_3 - t_2 t_3 + t_1 t_4 + t_2 t_4 + 2\, t_3 t_4\leq0
\end{array}
\eeq
The intersection of the half-spaces defined by the above inequalities with the K\"ahler cone, $t_i >0$, is non-empty, indicating the existence of supersymmetric vacua for generic (small) VEVs of the singlet fields and everywhere in a neighbourhood of $t_1=t_2=t_3=t_4$ in K\"ahler moduli space. For such generic VEVs, all the additional $U(1)$~symmetries are spontaneously broken. In addition, as long as $\langle S_{{\bf e}_2-{\bf e}_5}\rangle\neq 0$, an $L\bar{H}$ term is induced from the last term in the superpotential~\eqref{eq:W}. For a sufficiently large VEVs $\langle S_{{\bf e}_2-{\bf e}_5}\rangle$, this removes the pair of Higgs doublets from the spectrum. A non-generic Higgs doublet pair which is massless only for special choices of moduli is a common feature in string standard models - a string theory manifestation of the $\mu$-problem~\footnote{However, it is interesting to note that there are examples in the standard model data base~\cite{Linebundles} where the Higgs doublet pair remains massless throughout moduli space.}.

For this reason, we will focus on the part of moduli space where $\langle S_{{\bf e}_2-{\bf e}_5}\rangle =0$, while all other singlet field VEVs can be non-zero. In this case, for a solution to the D-flat conditions, the K\"ahler moduli should still satisfy the inequalities~\eqref{eq:Dterm2}, except for the fourth one which has to be replaced by the equality $-2\, t_1 t_2 + 2\, t_3 t_4=0$. A solution to these conditions exists for generic choices of all singlet VEVs (keeping $\langle S_{{\bf e}_2-{\bf e}_5}\rangle =0$) on a co-dimension one locus in K\"ahler moduli space. On this locus, we keep a light pair of Higgs doublets and only three of the four $U(1)$ symmetries are spontaneously broken, while one linear combination, denoted by $U_X(1)$ and specified by the direction $(1,1,1,1,-4)$ remains unbroken. We will now study the model at this locus in the moduli space in more detail.

\section{The $B-L$ model}\label{sec:NonAbelianLocus}
For 
$\langle S_{\mathbf{e}_2-\mathbf{e}_5} \rangle =0$ but otherwise generic,
non-zero singlet field VEVs, the low-energy theory has a symmetry group $G_{\text{SM}}\times U_X(1)$, where the group 
$U_X(1)$ is global, 
and the matter spectrum is given by
\begin{equation}\label{spectrum2}
\begin{array}{lll}
 3\,Q_{-1}~~~~~& 3\,u_{-1}~~~~~& 3\,e_{-1} \\[2pt]
 3\,L_{3}~~~~~& 3\,d_{3}~~~~~& 3\,\nu^R_{-5}\\[2pt]
 H_{-2}& \bar H_{2} & 9\,S_0
\end{array}
\end{equation}
 Here, the subscript denotes the $U_X(1)$ charge. The $U_X(1)$ symmetry is special in several ways. Firstly, combined with the hypercharge $Y$ as 
\beq
B -L \ =\ -\frac{1}{5}\, X\, +\, \frac{2}{5}\, Y \ .
\label{e1}
\vspace{-4pt}
\eeq
it leads to a $B-L$ symmetry of the model. Secondly, the spectrum contains three right handed neutrinos (which are identified with three of the singlet fields) with the correct $U_X(1)$ charge to render this additional symmetry non-anomalous. Note that the associated gauge boson is still St\"uckelberg heavy.

In Refs.~\cite{Buchbinder:2013dna, Buchbinder:2014qda} this model has been obtained both by continuation along flat directions from the model at the enhanced symmetry locus and by a direct string construction. As it stands, the model retains no memory of the additional three $U(1)$ symmetries present at the enhanced symmetry locus and is, therefore, much less restrictive. For example, the $U_X(1)$ symmetry allows for generic Yukawa couplings in contrast with the restricted superpotential~\eqref{eq:W}. However, we know that the superpotential at the enhanced symmetry locus is restricted as in Eq.~\eqref{eq:W} including all possible terms with singlet insertions. Hence, no other terms will be generated for non-zero singlet VEVs and we conclude that the superpotential retains its form~\eqref{eq:W}, despite the absence of symmetries in the $B-L$ model to explain this specific structure.

A similar situation arises with regard to proton stability. Dimension four operators, such as $u\,d\,d$, $Q\,L\,d$ and $e\,L\,L$ are forbidden by the $U_X(1)$ symmetry or, equivalently, by the $B-L$ symmetry. However, the global $U_X(1)$ symmetry does not forbid the dimension five operators $Q\,Q\,Q\,L$ and $u\,u\,d\,e$ which are known to induce fast proton decay. However, we know from Eq.~\eqref{eq:W} that all proton decay operators, including those with singlet insertions, are forbidden at the enhanced symmetry locus. Hence, the dimension five operators must be absent for the $B-L$ model even though they are not forbidden by a symmetry of the model.

Finally, we would like to discuss the possibility of discrete remnants from the spontaneous breaking of the three $U(1)$ symmetries which might explain the absence of these operators. The spontaneous breaking is induced by VEVs for the fields $S_{\mathbf{e}_2-\mathbf{e}_1}, S_{\mathbf{e}_4-\mathbf{e}_1}, S_{\mathbf{e}_2-\mathbf{e}_3}, S_{\mathbf{e}_4-\mathbf{e}_3}$. The $S(U(1)^5)$ charges indicated carry the correct integral normalization and they are all $\pm 1$. From these properties it can be shown that there are, in fact, no discrete remnants left over.

\section{Conclusion}\label{sec:Conclusion}

In this letter, we have studied a heterotic standard model in the context of its four-dimensional effective theory. Generically, this model has a standard model gauge group plus a $U_{B-L}(1)$ symmetry, an MSSM spectrum, including three right-handed neutrinos, and a number of singlet fields. In this model, dimension five operators which induce proton decay are allowed by the symmetries, indicating a possible phenomenological problem. However, we have shown that there is a specific locus in the moduli space where the symmetry enhances by three additional $U(1)$ symmetries. These additional $U(1)$ symmetries forbid the dangerous dimension five operators as well as all their possible variants with singlet insertions. As a result, these operators are absent throughout the moduli space and the $B-L$ model is safe from fast proton decay.

Our main point is that knowledge of the full moduli space of a model -- and loci of enhanced symmetry in particular -- can lead to phenomenologically important constraints on the model and can rule out couplings which seem otherwise allowed. Given that coupling constants, specifically those for higher-dimensional operators, are not easy to calculate directly from string theory this can lead to valuable information about the structure of the low-energy theory.\\[8mm]
{\bf\large Acknowledgements.}
We would like to thank Graham Ross for useful discussions and suggestions. The work of  EIB~is supported by the ARC Future Fellowship FT120100466. EIB would like to thank the Oxford University Physics Department where part of this work has been done, for warm hospitality.  AC's work is supported by the STFC. AL~is partially supported by the EPSRC network grant EP/l02784X/1 and by the STFC consolidated grant~ST/L000474/1.  


%
%

\end{document}